\documentclass[bibyear,usenatbib]{aa}  

\usepackage{graphicx}
\usepackage{txfonts}
\usepackage{hyperref}
\usepackage{lscape}
\usepackage{newtxtext,newtxmath}
\usepackage[T1]{fontenc}
\usepackage{ae,aecompl}
\usepackage{graphicx}	
\usepackage{amsmath}	
\usepackage{amssymb}	
\usepackage{rotating}
\usepackage{pdflscape}
\usepackage{natbib}
\usepackage{url}

\newlength\longest

\newcommand{\lumcgs}{ergs~s$^{-1}$}
\begin{document}

   \title{A \textit{NuSTAR} census of coronal parameters in Seyfert galaxies}

   \subtitle{}

   \author{A. Tortosa
          \inst{1}\thanks{E-mail: tortosa@fis.uniroma3.it}
          \and
           S. Bianchi\inst{1}
           \and
          A. Marinucci\inst{1}
          \and
          G. Matt\inst{1}
          \and
          P. O. Petrucci\inst{2}
          }

   \institute{Dipartimento di Matematica e Fisica, Universit\'a degli Studi Roma Tre, via della Vasca Navale 84, 00146 Roma, Italy.
   \and
   Univ. Grenoble Alpes, CNRS, IPAG, F-38000 Grenoble, France.\\}
   \date{Received 11, 29, 2017 ; accepted 12, 01, 2018}

 
  \abstract
   {We discuss the results on the hot corona parameters of Active Galactic Nuclei that have been recently measured with \textit{NuSTAR}. The values taken from the literature of a sample of nineteen bright Seyfert galaxies are analysed.}
   {The aim of this work is to look for correlations between coronal parameters, such as the photon index and cutoff energy (when a phenomenological model is adopted) or the optical depth and temperature (when a Comptonization model is used), with other parameters of the systems like the black hole mass or the Eddington ratio.}
   {We analysed the coronal parameters of the nineteen unobscured, bright Seyfert galaxies that are present in the \textit{Swift-BAT} 70 months catalogue and that have been observed by \textit{NuSTAR}, alone or simultaneously with others X-rays observatories such as \textit{Swift}, \textit{Suzaku} or \textit{XMM-Newton}.}
   {We found an anti-correlation with a significance level $>98$\% between the coronal optical depth and the coronal temperature of our sample. On the other hand, no correlation between the above parameters and the black hole mass, the accretion rate and the intrinsic spectral slope of the sources is found.}
   {}


   \maketitle
%

\section{Introduction}
\label{sec:intro}
The X-ray continuum of Active Galactic Nuclei can be explained by thermal Comptonization of the soft UV radiation, produced by the inner accretion disk, by a plasma of relativistic electrons around the supermassive black hole, known as the corona \citep{haardtmaraschi93}. This continuum is reprocessed by cold neutral circumnuclear medium (e.g. the accretion disc or the molecular torus) and gives rise to a reflection bump at around 30 keV and a  iron K$\alpha$ line emission at around 6.4 keV (e.g., \citealt{matt91}). The presence of these features in the spectrum places constraints on the geometry of the X-ray emitting region and tells us that the hot plasma region is quite compact and likely situated close to the accretion disc.\par 
However the detailed geometry of the disc/corona system is still largely unknown, and indeed the size, location and shape of the corona are still matter of debate. It is not yet clear if the corona is very compact, as assumed in the lamp-post geometry (\citet{matt91}, \citet{henripetrucci97}, \citet{petruccihenri97} \citet{miniutti2004}), or more extended, filling the inner part of the accretion flow. It is also not yet clear if the coronal plasma is a continuous or a patchy medium \citep{petrucci2013}. \par
To solve the doubts raised above we need to study the broad-band X-ray spectrum of AGN in details, to disentangle all the complex spectral features in this energy range, to remove all the degeneracies between the primary continuum features and other physical observables in order to constraints the coronal parameters and to have an overview of the physics and the structure of the hot corona \citep{2016mar}.\par 
The simplest way to obtain a description of Comptonizing coronae is to measure the cut-off in the hard X-ray spectrum and the photon power-law index. The spectral cut-off is directly related to the Comptonizing electron temperature of the corona while the power-law index depends on the interplay between the electron temperature and the optical depth. These kind of measurements have been performed with hard X-ray satellites, such as \textit{BeppoSAX} (\citealt{dadina2007}, \citealt{perola2002}), \textit{INTEGRAL} (\citealt{panessa2011}; \citealt{derosa2012}; \citealt{molina2013}) and \textit{Swift} (\citealt{Ricci2017}). The cutoff values  ranged between 50 and 300 keV. The high energy cutoff values result to be correlated with the photon index of the primary emission. It is not clear, however, if this was a real correlation of an effect due to the degeneracies between parameters in those background-dominated observations.\par
The Nuclear Spectroscopic Telescope Array (\textit{NuSTAR}, \citealt{harrison2013}) is a hard X-ray observatory launched in June 2012. It has two coaligned X-ray optics which focus X-ray photons onto two independent shielded focal plane modules, namely FPMA and FPMB. Thanks to its focusing optics, it has a broad and high quality spectral coverage from 3 to 79 keV. Given these features, \textit{NuSTAR} is suitable for studying the hard X-ray spectra of AGN with high sensitivity, discriminating between the primary X-ray emission and the scattered or reflected component (i.e. radiation which interacts with circumnuclear gas and gets absorbed or Compton scattered). Alone, or with
simultaneous observations with other X-ray observatories operating below 10 keV, such as \textit{XMM-Newton,} \textit{Suzaku} and \textit{Swift}, it provided strong constraints on the coronal properties of many AGN (\citealt{brenneman14}, \citealt{balokovic15}, \citealt{fabian15} \citealt{matt05}, \citealt{marinucci14a}, \citealt{2016mar}, \citealt{tortosa17}, \citealt{fabian17}).\par 
To better understand the complex environment of AGN, it is important to search for correlations between coronal parameters and other physical parameters, such as the black hole mass and the Eddington ratio. In this paper, we present the analysis of a small catalogue of AGN built up by choosing the unobscured nearby, non-jetted (following the distinction made by \citealt{padovani2017}) Seyfert galaxies that have been observed by \textit{NuSTAR} (often in coordination with \textit{XMM-Newton}, \textit{Suzaku} or \textit{Swift}). We took from the literature the values of the coronal parameters of this sample of AGN. The aim of the paper is to list and discuss these values and look for possible correlations. 


\section{The sample}
The primary X-ray emission is characterized by a power-law spectral shape extending to energies determined by the electron temperature. The power-law often shows a cutoff at high energies. Both the energy of the cutoff and the photon index are related to the temperature and the optical depth of the corona. Comptonization models imply that the cutoff energies are 2-3 times the temperature of the corona (\citep{petrucci2000, petrucci2001}). To investigate the shape of the spectrum it is important to take into account the reprocessed emission of the circumnuclear environment in this energy range, such as reflection from accretion disc and distant material. Typical X-ray features of the cold circumnuclear material include intense iron K alpha line at 6.4 keV and the associated Compton reflection peaking at 30 keV. \par 
Unlike the previous hard X-ray observatories, which are background dominated for almost all AGN, \textit{NuSTAR} is the first focusing hard X-ray telescope on orbit, 100 times more sensitive in the 10-79 keV band compared to previous observatories working in the same energy band. The focusing capability implies a very low background, and the observation of bright AGN are source-dominated. \textit{NuSTAR} data can, therefore, provide strong and robust constraints on the high-energy cutoff, allowing to study AGN at high energies with high precision and with unprecedented accuracy. Thanks to \textit{NuSTAR} observations in collaborations with other X-ray satellites such as \textit{XMM-Newton} and \textit{Swift}, in the last few years several cutoff energies have been measured with very high precision. \par
We built the catalogue by choosing the unobscured ($N_H \leq 6 \times 10^{22}$cm$^{-2}$) nearby brightest Seyfert galaxies that are present in the \textit{Swift-BAT} 70 months catalogue and that have been observed by \textit{NuSTAR} alone or simultaneously with other X-rays observatories, such as \textit{Swift}, \textit{Suzaku} or \textit{XMM-Newton}. We selected only unobscured, or moderately obscured AGN to have a clear view of
the primary emission component. 
Other objects for which the cutoff energy had been left fixed in the spectral analysis are not included (1H0707-495 for instance), since they need a more intensive study on this issue.The list and the characteristics of all the sources can be found in Table \ref{sample} and in Appendix \ref{list}. \par
\begin{figure}
\centering
\includegraphics[width=0.9\columnwidth]{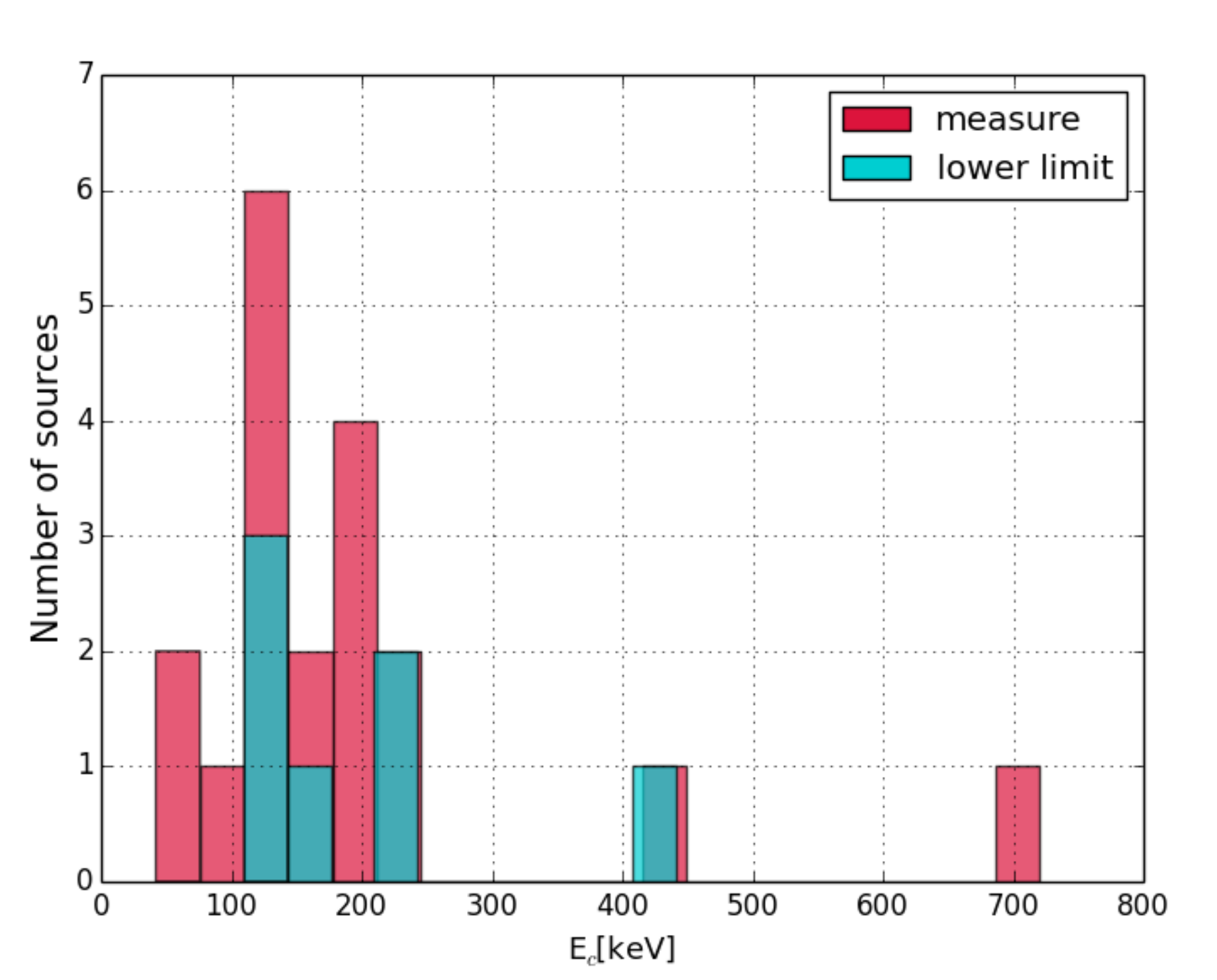}
\caption{The histogram of the distribution of the high-energy cutoff of the sample when both measures (red) and lower limits (blue) are considered.}
 \label{fig:histo}
\end{figure}

 
The final sample comprises nineteen objects, twelve with a measurement of the cutoff energy and seven having only a lower limit. The distribution of high-energy cutoff measurements from the sample is shown in Figure \ref{fig:histo}.

 \begin{table}
\caption{Correlations factor, $\rho$ and Null hypothesis probability, h$_0$.}             
\label{table:correlations}      
\centering          
\begin{tabular}{ccccc}  
\hline\hline   
X & Y & $\rho$ & h$_0$&geometry\\
\hline
$\Gamma$ & E$_c$& 0.18&0.47&-\\
 log(M$_{\rm bh}$/M$_\odot$) & E$_c$ & -0.11&0.61&-\\
 L$_{\rm bol}/$L$_{\rm Edd}$  & E$_c$&-0.14&0.56&-\\
 $\tau$ & kT$_e$ & -0.88 & 0.004 & slab\\
 $\tau$ & kT$_e$ & -0.63 & 0.02 & sphere\\
log(M$_{\rm bh}$/M$_\odot$) & $\tau$&  -0.22 & 0.63 &slab\\
 log(M$_{\rm bh}$/M$_\odot$) & $\tau$& -0.26 & 0.46 & sphere\\
 L$_{\rm bol}/$L$_{\rm Edd}$ & $\tau$ &0.49 & 0.27 &slab\\
 L$_{\rm bol}/$L$_{\rm Edd}$ & $\tau$ & 0.38 & 0.28 &sphere\\
 log(M$_{\rm bh}$/M$_\odot$) &  kT$_e$& 0.20 & 0.64 & slab\\
 log(M$_{\rm bh}$/M$_\odot$) &  kT$_e$& 0.18 & 0.47 & sphere\\
 L$_{\rm bol}/$L$_{\rm Edd}$ &  kT$_e$ & -0.37 & 0.41 & slab\\
 L$_{\rm bol}/$L$_{\rm Edd}$ & kT$_e$& -0.36 & 0.32 & sphere\\
\hline
\hline                  
\end{tabular}
\end{table}
 \subsection{Black Hole mass measurements}
Some of the selected sources had more than one literature value for the mass of the central black hole. One of the most reliable and direct way to measure the mass of a super massive black hole residing in the nucleus of an active galaxy is reverberation mapping (RM, \citealt{blandford1982}; \citealt{peterson1993}). We decided to use the RM mass values, for the sources that have one (IC4329A, 3C390.3, Ark 564, Ark 120, Mrk 335, Fairall 9, Mrk 766, PG 1211+143 \citealt{peterson04}; NGC 6814 \citealt{pancoast2014, pancoast2015}). For the sources without a RM measurement we used mass values derived from virial mass methods, such as the single-epoch method (SE). These methods start from the relation between the size and the luminosity of the broad line region (R-L relation), to derive the broad line region size through a single measure of the optical continuum luminosity and, combining this information with the width of a broad line, it is possible to build a relation with the black hole mass \citep{vestergaard2002,vestergaard2006}. 
  \begin{landscape}
 \begin{table}
 \caption{Spectral parameters, masses, luminosity and accretion rates of the sources of the selected sample. Accretion rates are computed using the $L_{x}$ in the 2-10 keV energy band using the bolometric correction of \citet{marconi2004}. Luminosity is in unit of $10^{44}$ \lumcgs. Flux is in unit of $10^{-11}$ erg cm$^{-2}$s $^{-1}$. The bottom part of the table is for objects with lower values of the high energy cutoff.}
     \label{sample}
    \centering   
    \renewcommand\arraystretch{1.5}
\begin{tabular}{ccccccccccccc}
Source &Ref. &  $\Gamma$  & $E_c$  & log(M$_{\rm bh}$/M$_\odot$)& Ref. & L$_{\rm bol}/$L$_{\rm Edd}$ & L$_{2-10 keV}$ &F$_{\rm 2-10 keV}$  & kT$_e$& $\tau$& geom.&model\\
\hline
&& &[keV]& && & \lumcgs & erg cm$^{-2}$s $^{-1}$ &[keV]& & &\\
 \hline
NGC 5506 & 1 & $1.91 \pm 0.03$ & $720^{+130}_{-190}$ & $8.0 \pm 0.2$& (A) & $0.006$ & $0.053$ &$6.2$ & $440^{+230}_{-250}$ & $0.02^{+0.2}_{-0.01}$&slab&\textsc{compTT}\\
& & & & & &&& & $440^{+230}_{-250}$ & $0.09^{+0.2}_{-0.01}$ & sphere&\textsc{compTT}\\
MCG -05-23-16 &2 & $1.85 \pm 0.01$ & $170 \pm 5$ & $7.7 \pm 0.2 $&(B) & $0.058$ & $0.18$ &$10.4$ & $30 \pm 2$& $1.2 \pm 0.1$& slab&\textsc{compTT} \\
& & & & && & && $25 \pm 2$ & $3.5 \pm 0.02$ &sphere&\textsc{compTT}\\
SWIFT J2127.4 &3-4 & $2.08 \pm 0.01$ & $180^{+75}_{-40}$ & $7.2 \pm 0.2$&(J) & $0.136$ & $0.14$ & $2.9$ &$70^{+40}_{-30}$& $0.5^{+0.3}_{-0.2}$& slab&\textsc{compTT}\\
& & & & & &&& & $50^{+30}_{-25}$& $1.4^{+1.0}_{-0.7}$& sphere&\textsc{compTT}\\
IC4329A &5-6 & $1.73 \pm 0.01$ & $185 \pm 15$ & $8.08\pm 0.3$&(N) & $0.125$ & $0.56$ &$12.0$ & $37 \pm7 $ & $1.3 \pm 0.1$ &slab&\textsc{compTT}  \\
 & & & & && & && $33 \pm 6$ & $3.4 \pm 0.5$ & sphere&\textsc{compTT}\\
3C390.3&7& $1.70 \pm 0.01$   &$120 \pm 20$   &   $8.4 \pm 0.4$ &(H) & $0.241$ & $1.81$ &$4.03$ & $40 \pm 20$&$3.3^{+1.3}_{-2.8}$&sphere&\textsc{compTT}\\
3C382& 8 & $1.68\pm 0.03$ & $215^{+150}_{-60}$ & $9.2 \pm 0.5$&(D) & $0.072$ & $2.34$& $2.9$&$330 \pm 30$&$0.2 \pm 0.02$& slab&\textsc{compTT} \\
GRS 1734-292&9& $1.65 \pm 0.05$ & $53 \pm 10$ & $8.5 \pm 0.1$&(L) & $0.036$ & $0.056$ & $2.9$&$12 \pm 1$&$2.9 \pm 0.2$ &slab&\textsc{compTT}\\ 
 & & & & & & &&& $12^{+1.7}_{-1.2}$ & $6.3 \pm 0.3$ &sphere&\textsc{compTT}\\
NGC 6814&10 & $1.71 \pm 0.04$ & $135^{+70}_{-35}$ & $7.0 \pm0.1$&(C) & $0.003$ & $0.021$ & $0.2$ & $45^{+100}_{-20}$ & $2.5^{\dagger} \pm 0.5$ & sphere&\textsc{nthcomp}\\
MCG +8-11-11&10 &  $1.77\pm0.04$ & $175^{+110}_{-50}$ & $7.2 \pm 0.2$&(E) & $0.754$ & $0.51$& $5.6$ & $60^{+110}_{-30}$&$1.9^{\dagger} \pm 0.4$ &sphere&\textsc{nthcomp}\\
Ark 564&11 & $2.27 \pm 0.08$ & $42 \pm 3$ &  $6.8 \pm 0.5$& (H) & $1.313$& $0.39$ &-& $ 15 \pm 2$ & $2.7^{\dagger} \pm 0.2$ & sphere&\textsc{nthcomp}\\
PG 1247+267&12-13&$2.35 \pm 0.09$& $90^{+130}_{-35}$ & $8.9 \pm 0.2$&(M) & $0.024$ & $0.79$ &$0.05$&$46^{+60}_{-20}$& $1.4^{\dagger} \pm 0.3$&sphere&\textsc{nthcomp}\\ 
Ark 120 &14-15& $1.87 \pm 0.02$ & $180^{+80}_{-40}$ & $8.2 \pm 0.1$&(H) & $0.085$ & $0.92$&$2.3$ &-&-&-&-\\
\hline
NGC 7213&16& $1.84 \pm 0.03$ & $>140$ & $8.0 \pm 0.2$&(G) & $0.001$ & $0.012$& $1.3$&$230^{+70}_{-250}$ & $0.2\pm 0.1$&sphere&\textsc{compPS}\\
MCG 6-30-15& 17-18& $2.06 \pm 0.01 $ & $>110$ & $6.4 \pm 0.1$ &(E) & $0.238$ & $0.056$ & $5.5$ &-&-&-&-\\
NGC 2110 & 19 & $1.65 \pm 0.03$ & $>210$ & $8.3 \pm 0.2$ &(K) & $0.016$&$0.35$ &$12.5$ & $190 \pm 130$ & $0.2\pm 0.1$&slab&\textsc{compTT} \\
Mrk 335 & 21-22 & $2.14\pm 0.03$ & $>174$ & $7.2 \pm 0.1$&(H) & $0.284$ & $0.18$ & $1.9$ &- &-&-&-\\
Fairall 9 & 20 & $1.95 \pm 0.02$ & $>242$ & $8.1 \pm 0.7$& (H)& $0.054$ & $0.60$&$2.9$&-&-&-&- \\
Mrk 766 & 17-23-24 & $2.22\pm 0.03$& $>441$ & $6.3 \pm 0.1$& (I) & $1.254$ &$0.046$ &$1.4$& - &- &-&-\\
PG 1211+143 & 26& $2.51 \pm 0.2$ & $>124$ &$8.2 \pm 0.2 $&(H) &$0.047$ & $0.35$&$1.0$&-&-&-&-\\
    \end{tabular}
    \tablebib{1. \citet{matt05}, 2. \citet{balokovic15}, 3.\citet{marinucci14a}, 4.\citet{malizia08}, 5. \citet{bianchi09}, 6. \citet{brenneman14}, 7. \citet{lohfink15}, 8. \citet{ballantyne14}, 9. \citet{tortosa17}, 10. \citet{tortosa17sub}, 11. \citet{kara2017}, 12. \citet{lanzuisi16}, 13. \citet{trevese14}, 14. \citet{matt14}, 15. \citet{porquet2017}, 16. \citet{ursini15}, 17. \citet{marinucci14b}, 18. \citet{emmanoulopoulos14}, 19. \citet{marinucci15},  20. \citet{lohfink16}, 21. \citet{parker2014}, 22. \citet{bianchi2001}, 23. \citet{buisson17}, 24. \citet{risaliti2011}, 25. \citet{zoghbi15}, 26. \citet{Lawson1997}.\\
\textbf{Mass References:} A. \citet{papadakis2004}, B. \citet{onori2017}, C. \citet{pancoast2014, pancoast2015},D. \citet{winter09},  E.\citet{bian2003}, F. \citet{zhang06}, G. \citet{woo2002},  H. \citet{peterson04},  I. \citet{bentz10},  J\citet{halpern2006},  K. \citet{moran07}, L. \citet{tortosa17}, M. \citet{trevese14}, N \citet{delacalle}.}
\begin{flushleft}
$\dagger$ Estimated parameter.\\ 
\end{flushleft}
\end{table}
\end{landscape}
 One of the most used R-L relations based on H$\beta$ RM measurements is \citep{bentz2009}:
\begin{equation}
\log{\frac{R}{lightdays}}=-2.13+0.519\log{\frac{\lambda L_{\lambda}(5100 \AA)}{\rm erg \rm s^{-1}}}
\end{equation}
In the case of NGC 5506 the central stellar velocity dispersion ($\approx 180$km s$^{-1}$) (\citealt{oliva1999}, \citealt{papadakis2004}) and the width of the [OIII] line \citep{boroson2003} give a black hole mass $\sim 10^8$M$_{\odot}$, and we decided to use this value. \par 
We assumed a 20\% uncertainty for black hole mass estimates not inferred from reverberation.
\section{The fitting process}
\label{fitprocess}
The aim of this work is, as said before, to look for correlations between the spectral parameters, such as the cutoff value and physical parameters.\par
The goodness of the correlation is given by the Spearman's rank correlation coefficient or Spearman's $\rho$.
The Spearman correlation coefficient is defined as the Pearson correlation coefficient between the ranked variables. The sign of the Spearman correlation indicates the direction of the association between X (the independent variable) and Y (the dependent variable). A Spearman correlation close to zero indicates that there is no tendency for Y to either increase or decrease when X increases. When X and Y are perfectly monotonically related, the Spearman correlation coefficient becomes $1.0$ (or $-1.0$ for anti-correlation).
In the fitting process also the "null hypothesis" is given. The null hypothesis, denoted by $h_0$, is the probability that sample observations result purely from chance. A small $h_0$ value indicates a significant correlation.
The fits are made using the Interactive Data Language (IDL \footnote{\url{www.harrisgeospatial.com/SoftwareTechnology/IDL.aspx}}) programming language. We fitted the parameters with a simple linear relation in logarithmic scale:
\begin{equation}
\log(y)=\textbf{a}\log(x)+\textbf{b}
\label{eq:fit}
\end{equation}
The fits are made with a Monte Carlo method which repeated the fit procedure by random sampling the values between a minimum value and a maximum value, which are identified respectively with the lower and the upper extreme of the errors of the measure, \citep{bianchi2009}. 
\subsection{Spectral Parameters}
\begin{figure}
\centering
\includegraphics[width=0.9\columnwidth]{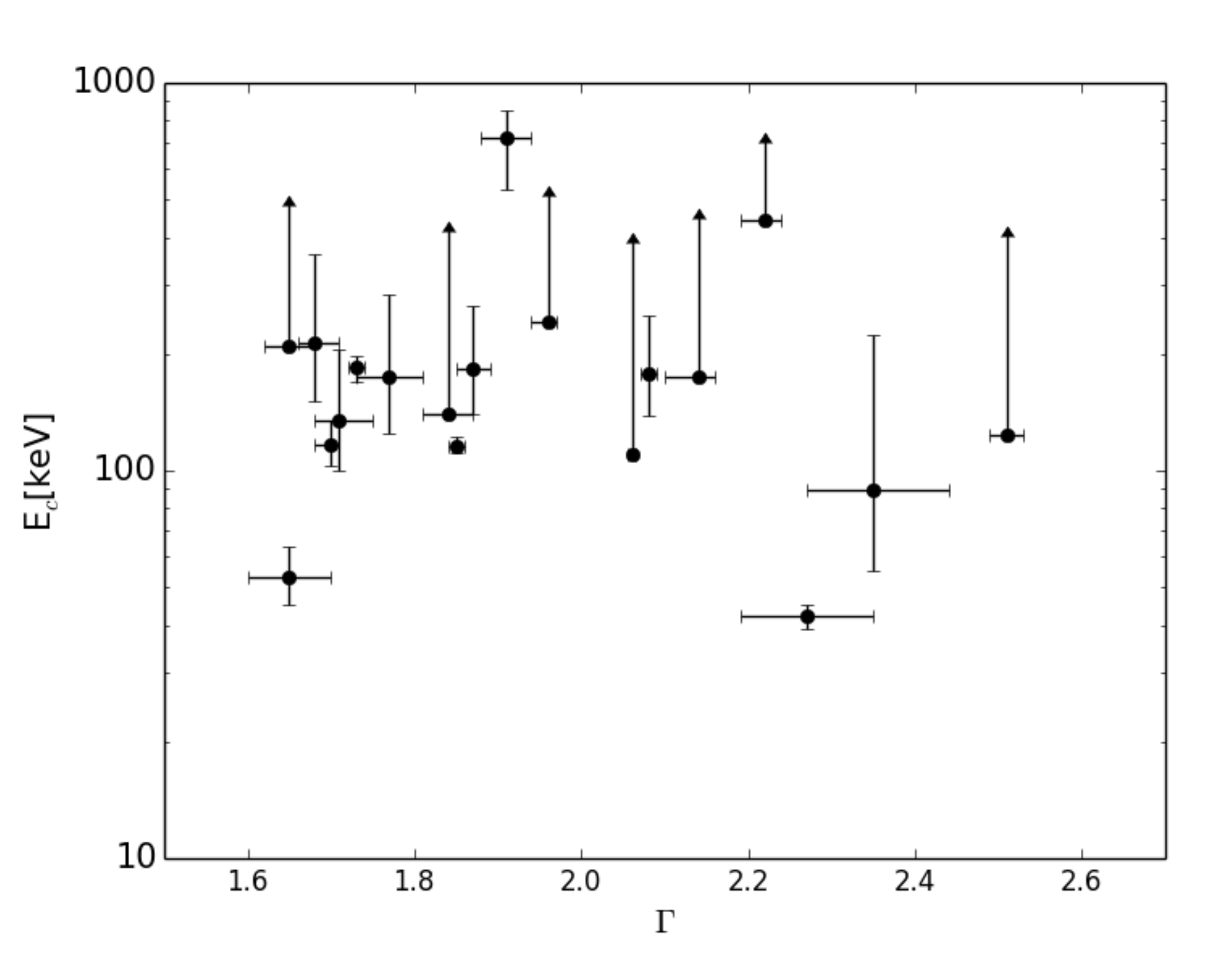}
\caption{Plot of the high-energy cutoff versus the photon index of the sample.}
 \label{fig:g_vs_ec}
\end{figure}

We started by looking for a correlation between the photon index $\Gamma$ and the high-energy cutoff with the relation of Equation \ref{eq:fit}. As it can be seen in Figure \ref{fig:g_vs_ec}, no statistically significant correlation is found between this parameters, in contrast with what was found by previous satellites ( e.g., \citealt{perola2002}; \citealt{Ricci2017}). This is reassuring, suggesting that with \textit{NuSTAR}, and using a relatively large sample of well exposed sources with good measurements, the intrinsic degeneracy between these two parameters is significantly reduced.
The following step was to search for a linear correlation between the high energy cutoff and either the mass of the central black hole or the Eddington ratio.
\begin{figure}
\centering
\includegraphics[width=0.9\columnwidth]{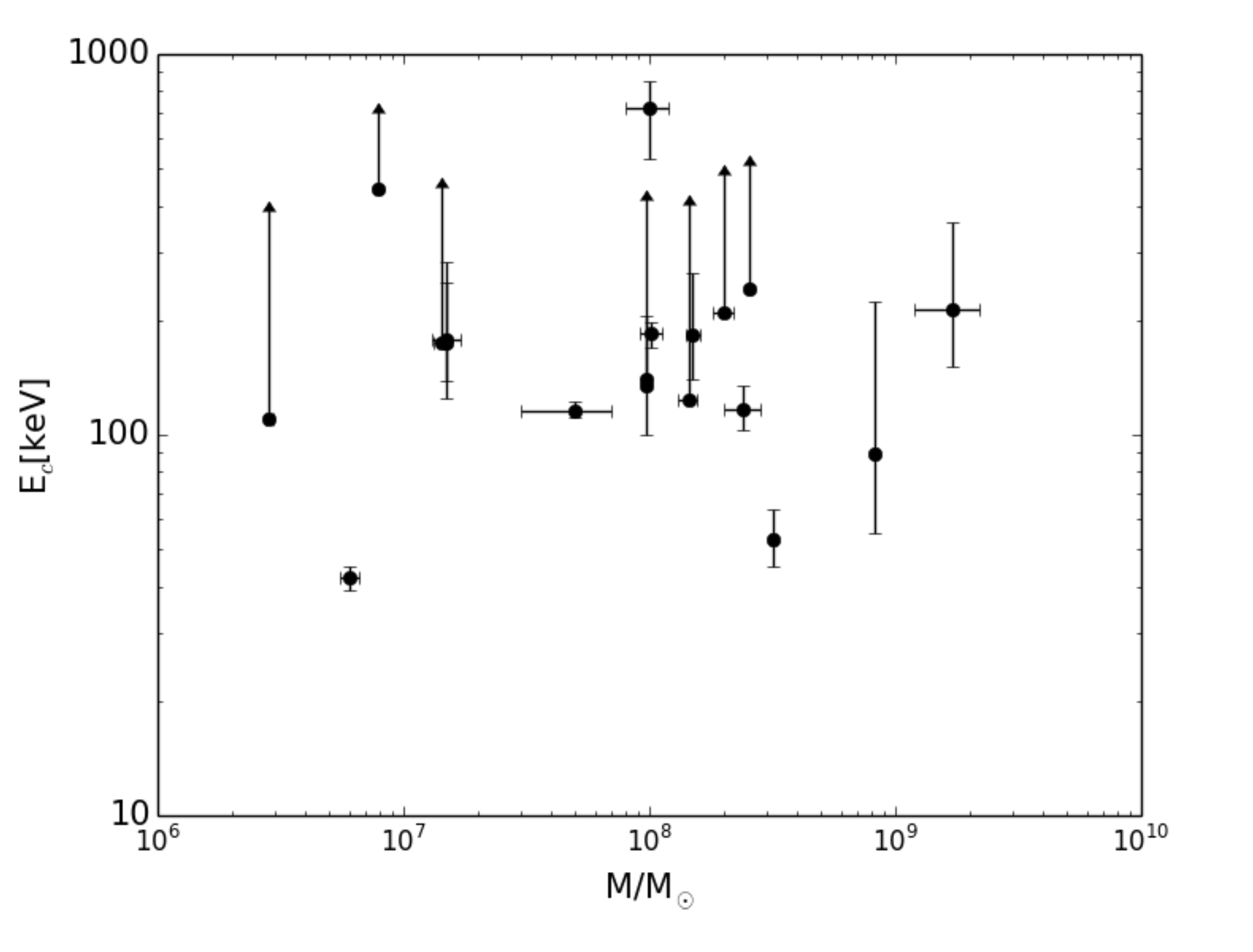}
\caption{Plot of the high energy cutoff versus the black hole mass of the sample}
 \label{fig:m_vs_e}
\end{figure}
\begin{figure}
\centering
\includegraphics[width=0.9\columnwidth]{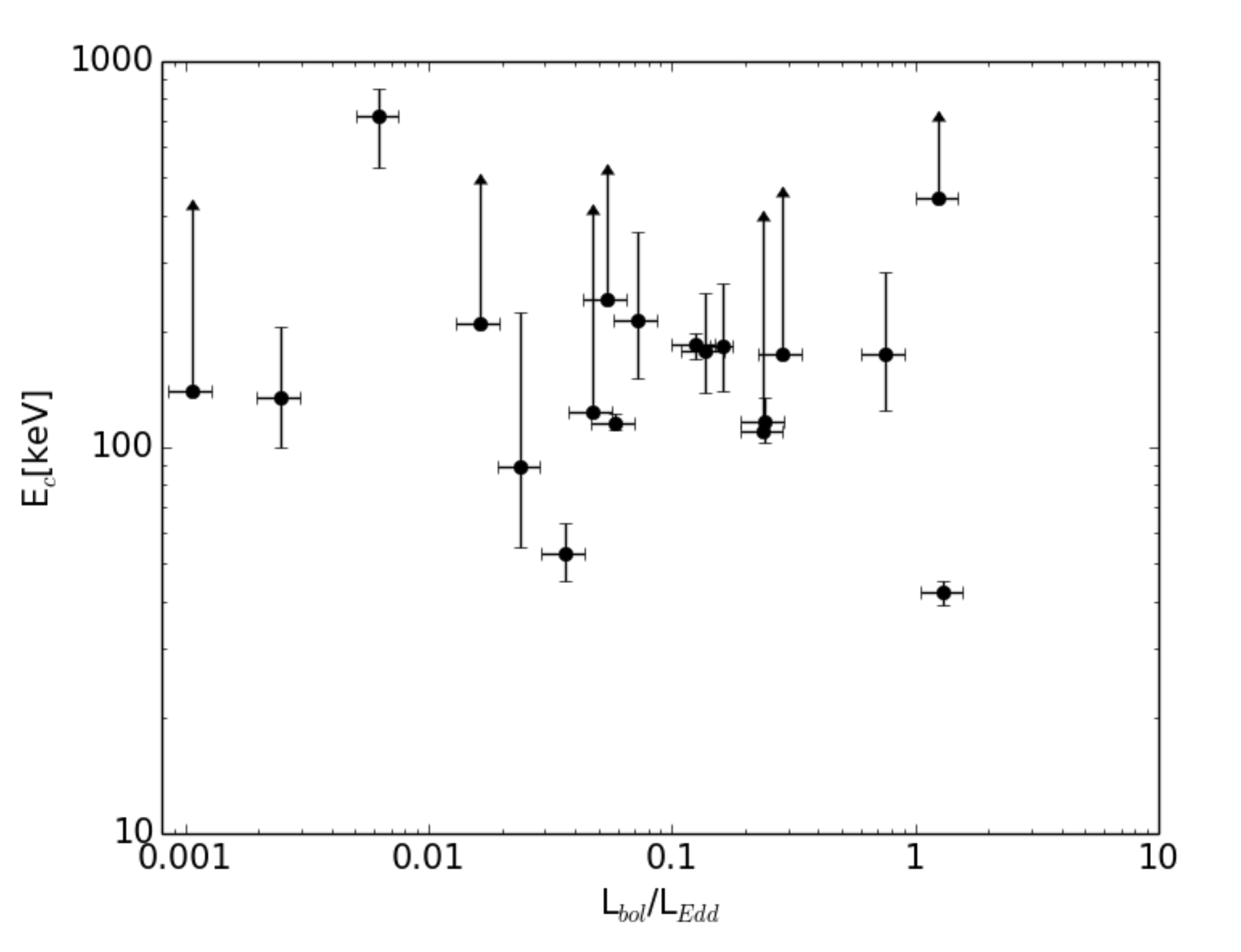}
\caption{Plot of the high energy cutoff versus Eddington ratio of the sample}
 \label{fig:e_vs_e}
\end{figure}
The Eddington ratio L$\_{\rm bol}/$L$_{\rm Edd}$ is computed using the $2-10$ keV absorption-corrected luminosity of the sourced to estimate the bolometric luminosity using the $2-10$ keV bolometric correction of \citealt{marconi2004}. Error bars on the Eddington ratio are derived from uncertainties on the black hole mass and $2-10$ keV luminosity measurements. All values are reported in Table \ref{sample}. The Spearman's $\rho$ values, reported in Table \ref{table:correlations}, show that there is no significant correlation between the checked parameters. \par
\subsection{Physical Parameters}
\begin{figure*}
\centering
\includegraphics[width=0.9\columnwidth]{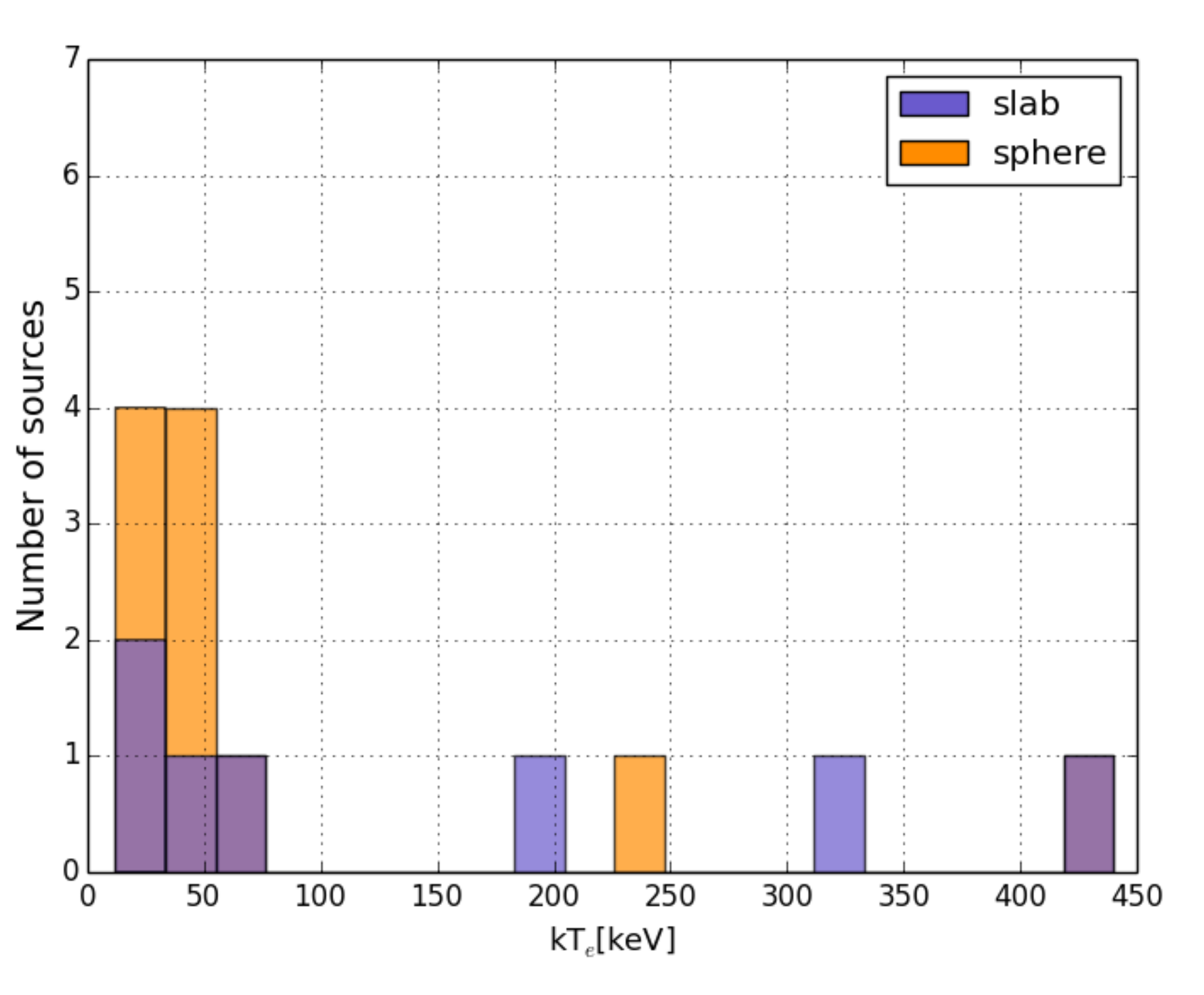}
\includegraphics[width=0.9\columnwidth]{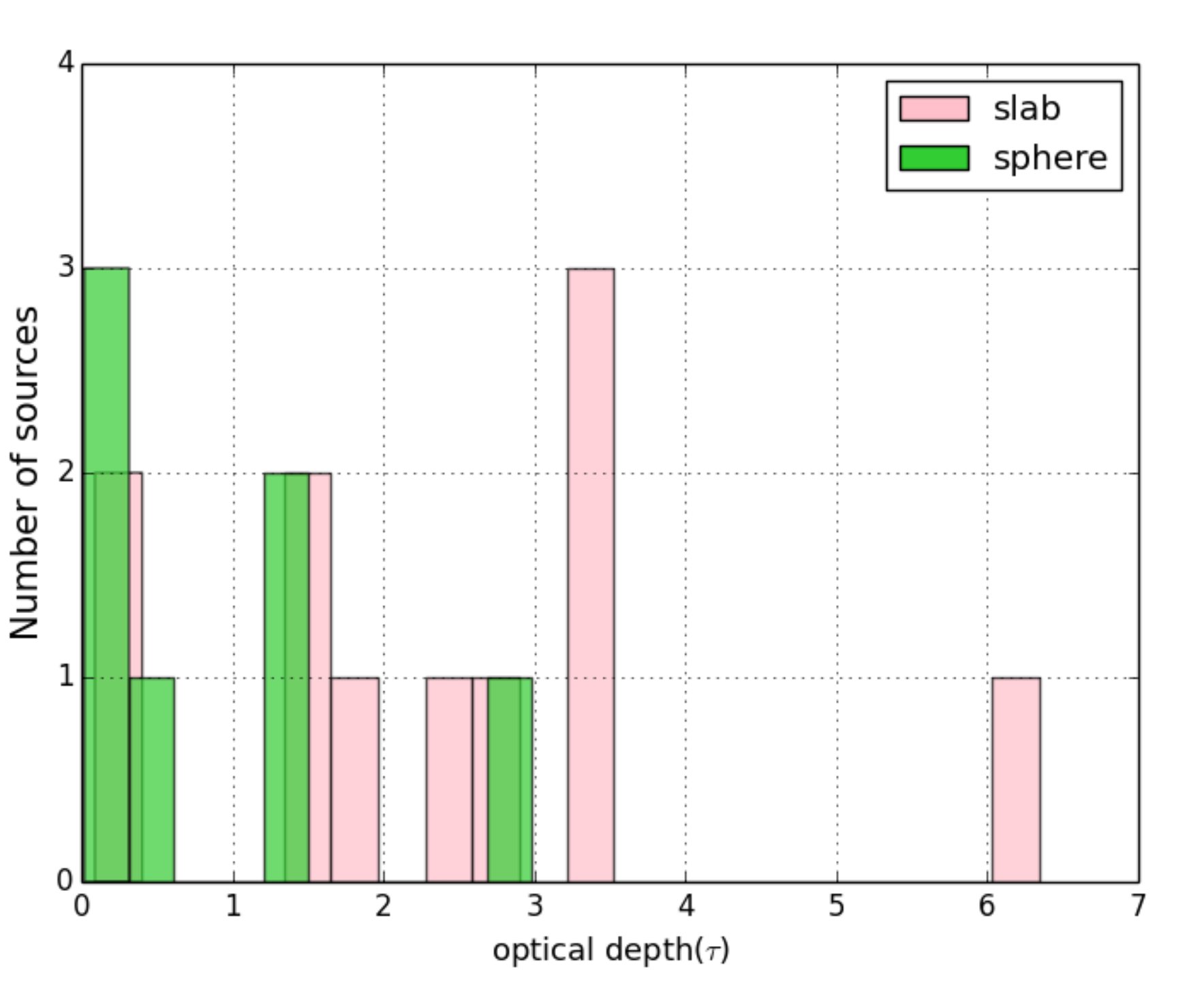}
\caption{Left panel: the histogram of the distribution of the coronal temperature values for the sources of the sample that have coronal temperature measurements. Right panel: the histogram of the distribution of the optical depth values for the sources of the sample that have a direct or extrapolated measurements of this parameter. Both slab and sphere geometry of the corona are considered.}
 \label{fig:histo_kt_t}
\end{figure*}

We consider the physical parameters that characterize the AGN coronae: the coronal temperature $kT_e$ and the optical depth. The distribution of these two values in our sample is shown in Figure \ref{fig:histo_kt_t}.
It should be noted that for some of the sources the optical depth parameter is not directly measured, since the model used (\textsc{nthcomp} in\textsc{xspec}) does not have the optical depth as free parameter. In these cases, the optical depth has been estimated using the relation from \citet{belodorov1999}:
\begin{equation}
\Gamma \approx \frac{9}{4} y^{-2/9}
\end{equation}
where $\Gamma$ is the photon index of the spectrum between 2 and 10 keV. The dependence from the optical depth is in the relativistic $y$-parameter:
\begin{equation}
y=4\left( \Theta_e + 4 \Theta_e^2\right) \tau(\tau+1)
\end{equation}
where $\Theta_e$ is the electron temperature normalized to the electron rest energy:
\begin{equation}
\Theta_e=\frac{kT_e}{m_ec^2} 
\label{theta}
\end{equation} 
We performed the fit for the two cases of slab and spherical geometry of the corona.\par
In Appendix \ref{reanalysis}, the results obtained with \textsc{nthcomp} are compared with those obtained with \textsc{compTT} for a few selected sources.
\subsubsection{Optical Depth vs Coronal Temperature}
\begin{figure}
\centering
\includegraphics[width=1.\columnwidth]{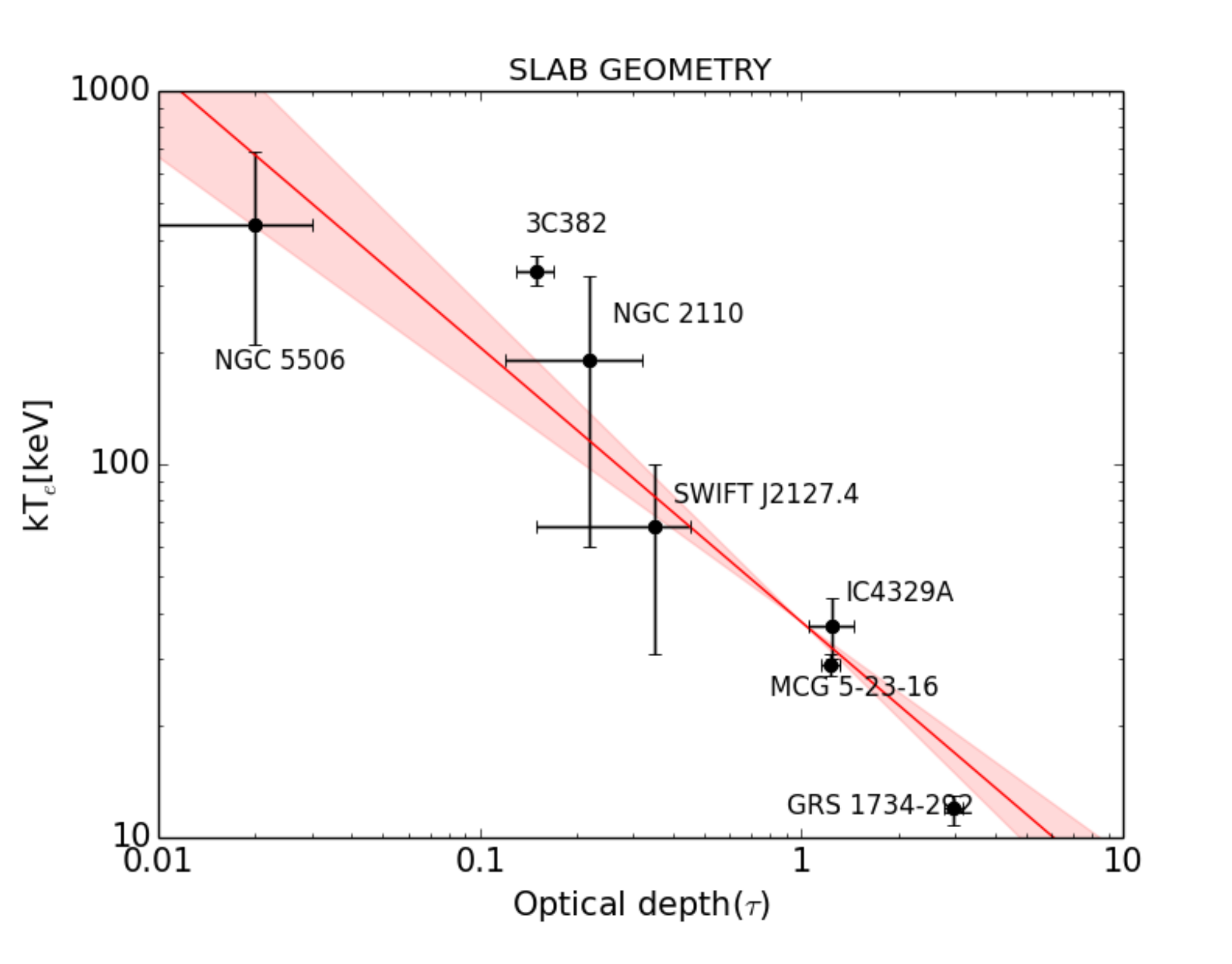}
\includegraphics[width=1.\columnwidth]{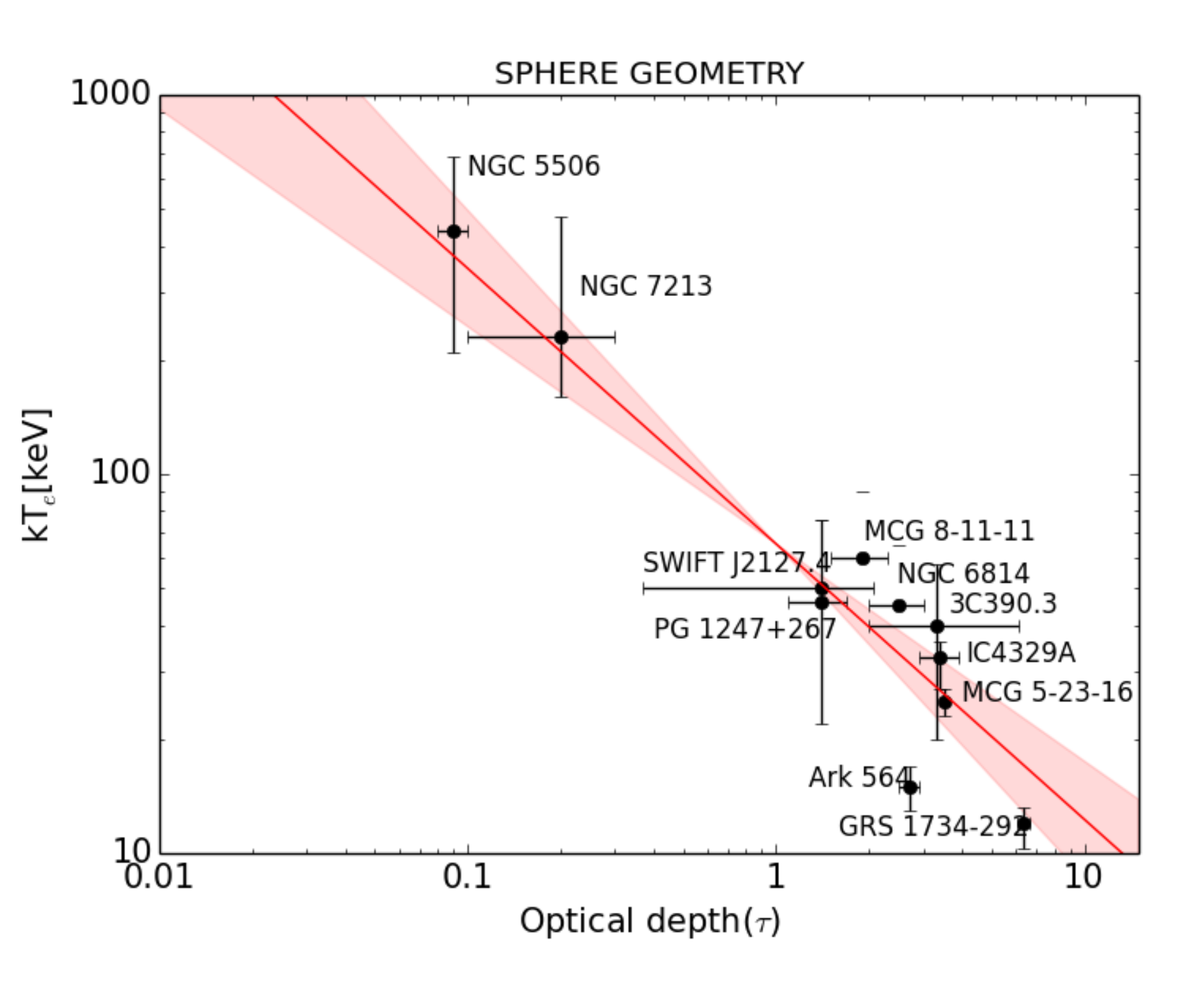}
\caption{Fit (red line) and error on the fitting relation (red shaded region) of the optical depth versus the electron coronal temperature in the case of a disc shape corona (top panel) and spherical corona (bottom panel). The fit is made using the relation \ref{eq:fit}.}
 \label{fig:kt_vs_t}
\end{figure}

The optical depth and coronal temperature appear to be extremely anti-correlated, the Spearman correlation factor for this fit is $\rho=-0.88$/$-0.63$ and the null hypothesis probability $h_0=0.004$/$0.03$, for the slab and spherical geometry respectively, see also Figure \ref{fig:kt_vs_t}. \par
Using the equation \ref{eq:fit} we found, in the case of a slab geometry, the following intercept and slope values for the fit:
 \begin{equation}
\textbf{a}=-0.7 \pm 0.1  \; ; \; \textbf{b}= 1.6 \pm 0.06
 \end{equation}
 The parameters of the linear regression in the case of the spherical geometry are:
  \begin{equation}
 \textbf{a}=-0.7 \pm 0.2 \; ; \; \textbf{b}= 1.8 \pm 0.1
 \end{equation}
 This is a very interesting result, but the physical interpretation is not straightforward. We will discuss this correlation in the following section.\par
We also searched for correlations between the above parameters (coronal optical depth and coronal temperature) and the central black hole of the AGN, and the Eddington Ratio in both coronal geometries, slab and spherical. We do not found statistically significant correlation any of the analysed cases (see Table \ref{table:correlations}). \par
\section{Discussion}
We found two relevant results from this analysis. The first one is the lack of correlation between the high-energy cutoff and the spectral photon index of the primary power law (see Figure \ref{fig:g_vs_ec}). \citeauthor{perola2002} in 2002 found a correlation between the high-energy cutoff and the photon index of the primary power law with a correlation coefficient equal to $0.88$, with E$_c$  increasing with $\Gamma$. The same correlation is found in the \textit{Swift-BAT} sample, in which \citet{Ricci2017} found that, when fitting the simulated \textit{Swift}/BAT spectra with a simple power-law model, the \textit{Swift}/BAT photon index increases when the cutoff decreases (see Figure 19 in their work).\par
Given that the two parameters are correlated in the fit procedure, this correlation may be an artifact due to any systematic error on one of the two parameters. Instead, we found no significant correlation between $\Gamma$ and E$_c$. The lack of correlation between these parameters means that there are no large systematics in the \textit{NuSTAR} measurements.\par
\begin{figure}
\centering
\includegraphics[width=1\columnwidth]{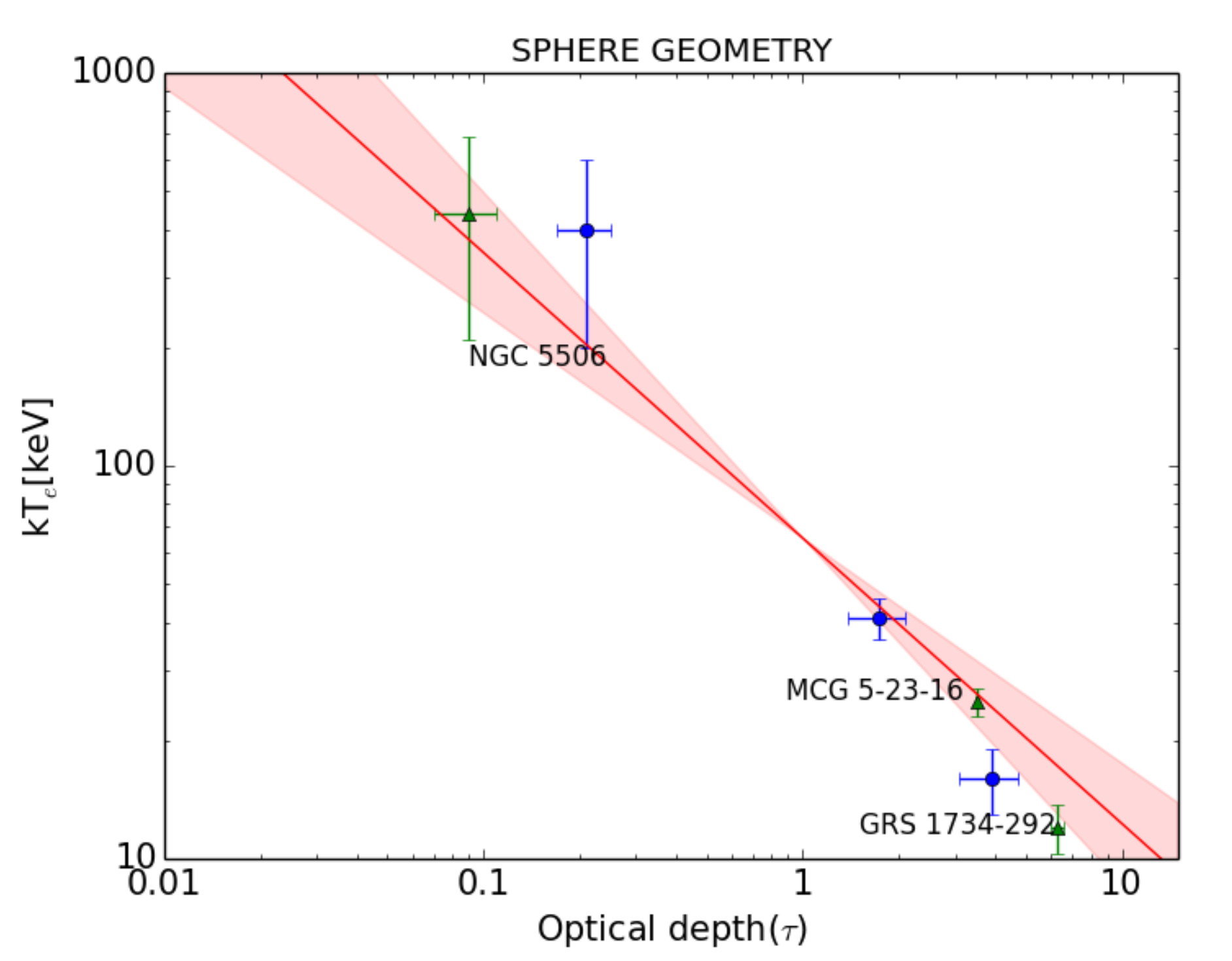}
\caption{Optical depth versus electron coronal temperature in the case of a spherical corona for GRS 1734-292, NGC 5506 and MCG -05-23-16 (blue circle) and the values obtained with our re-analysis (green triangle), see Appendix \ref{reanalysis}. We superimposed the fit (red line) and the error on the fitting relation (red shaded region) obtained with all the literature values for all the sources, as in lower panel of Figure \ref{fig:kt_vs_t}.}
 \label{fig:check}
\end{figure}

The second important result is the presence of a strong anticorrelation between the optical depth and the coronal temperature, both in the slab and in the spherical geometry.The interpretation of this anticorrelation is not trivial. Of course the values of the parameters are model dependent. We check the kT$_e$ and $\tau$ values in spherical geometry of the corona for some of the sources of the sample, in particular GRS 1734-292, NGC 5506 and MCG -05-23-16, which have very different values of temperature and optical depth, by re-analysing the \textit{NuSTAR} observations. The coronal temperature and the optical depth of the three sources listed above are all obtained with the \textsc{compTT} model \citep{titcomptt}. Instead we used the \textsc{nthcomp} model (\citealt{zdziarski1996} and \citealt{zycki1999}), see Appendix \ref{reanalysis}. We found different values for the two parameters of the three sources, but they still follow the anti-correlation (see Figure \ref{fig:check}). \par
Moreover, we note that the model used for the analysis of the different sources in the literature is not always the same. This excludes the fact that the correlation could be an artefact due to the use of the same model for the analysis. \par 
The $\tau$-kT$_e$ anti-correlation cannot be easily reconciled with a {\it fixed} disc-corona configuration in radiative balance. Indeed, such a configuration corresponds to a fixed cooling/heating ratio for the corona. In this case the corona temperature and optical depth have only to adjust themselves in order to ensure the constancy of this ratio. But there is no reason for kT$_e$ and/or $\tau$ to change. In other words, if 1) the disc-corona configuration of all the Seyfert galaxies is the same and 2) is in radiative balance, we would expect kT$_e$ and/or $\tau$ to cluster around the same values for all the objects of our sample.\par 
The observed correlation indicate that one (or both) of these hypotheses is wrong. The invalidation of the former (same disc-corona configuration) implies a geometrical variation of the accretion flow. It could be the variations of the transition radius R$_{tr}$ separating the inner corona and the outer disk or the variation of the height H of the corona above the disk, like in the lampost configuration. A smaller  R$_{tr}$/H would imply a larger cooling from the disk and then a smaller temperature (assuming the heating is the same). In this case the observed anti-correlation  would indicate that objects like NGC 5506 have a larger  R$_{tr}$/H than objects like GRS 1734-292.\par 
The invalidation of the radiative balance hypothesis, instead, could be due to, e.g., a variation of the fraction of thermal emission due to viscous dissipation ("intrinsic emission" from now on) with respect to the total disk emission, which also includes reprocessing of the coronal radiation. Indeed, for a fixed disk-corona geometry, the radiative balance will change if this proportion varies. If it increases, the cooling of the corona will increase and the temperature will decrease. In this case the observed anti-correlation would indicate that the disk intrinsic emission in objects like NGC 5506 has a smaller contribution to the total disk emission compared to objects like GRS 1734-292.\par
Note that for a pair-dominated corona, opposite behaviours are expected since an increase of the cooling (which is inversely proportional to the coronal optical depth, \citealt{haardtemaraschi91}) would correspond to an increase of the corona temperature and not a decrease \citep{ghisellinihaardt1994}. As a consequence, to explain the observed kT$_e - \tau$ anti-correlation, objects with large corona temperatures would have a smaller R$_{tr}$(H) or a larger contribution of the disk intrinsic emission than objects with low temperature. 

\section{Conclusions}
We have presented and discussed recent high-energy cutoff measurements in a sample of nineteen bright Seyfert galaxies observed by \textit{NuSTAR} in collaboration with other X-ray observatories operating below 10 keV, such as \textit{XMM-Newton,} \textit{Suzaku} and \textit{Swift}. The goal of the work is to look for correlation between spectral and physical parameters, to better understand the physics and the structure of AGN coronae.\par
This kind of analysis has been already done before the coming of \textit{NuSTAR} using cutoff energy measurements made up by hard X-ray satellites like \textit{Beppo-SAX}\citep{perola2002} and \textit{INTEGRAL} \citep{malizia2014}. Unlike \textit{NuSTAR}, these instruments are non-focusing, and therefore background dominated for AGN observations.\par
We searched for correlations between the high-energy cutoff and the photon index of the primary power-law, the mass of the central black hole and the Eddington ratio, i.e. L$_{\rm bol}/$L$_{\rm Edd}$. We did not found any statistically significant correlation between these parameters.\par 
Finally, we searched for correlations between the physical parameters which characterize the hot coronae of AGN, i.e. the temperature and the optical depth, and the mass of the central black hole and the Eddington ratio. No significant statistical correlation is found between these parameters. Instead, a significant anti-correlation is found between the optical depth and the coronal temperature fit. We found a Spearman correlation coefficient $\rho=-0.88$ in the case of a slab geometry of the corona and $-0.63$ in the case of a spherical corona. The null hypothesis probability, $\rho$, is equal to $0.004$ in the case of slab geometry and $0.02$ for the sphere geometry. 
The observed anti-correlation suggests a disk-corona configuration in radiative balance, but requires differences, from source to source, in either the disk-corona configuration or in the intrinsic disk emission.

\begin{acknowledgements}
      This work made use of data from the \textit{NuSTAR} mission, a project led by the California Institute of Technology, managed by the Jet Propulsion Laboratory, and funded by the National Aeronautics and Space Administration. We thank the \textit{NuSTAR} Operations, Software and Calibration teams for support with the execution and analysis of these observations. This research has made use of the\textit{ NuSTAR} Data Analysis Software (NuSTARDAS) jointly developed by the ASI Science Data Center (ASDC, Italy) and the California Institute of Technology (USA). AT, AM and GM acknowledge financial support from Italian Space Agency under grant  ASI/INAF I/037/12/0-011/13, SB under grant ASI-INAF I/037/12/P1. Part of this work is based on archival data, software or on-line services provided by the ASI SCIENCE DATA CENTER (ASDC).
AT, SB, AM and GM  acknowledge financial support from the European Union Seventh Framework Programme (FP7/2007-2013) under grant agreement n.312789. POP acknowledges financial support from the CNES and the CNRS/PNHE. EP and SB acknowledge financial contribution from the agreement ASI-INAF I/037/12/0. 
\end{acknowledgements}

%
   \bibliographystyle{aa} 
   \bibliography{biblio} 
   
\begin{appendix} 
\section{Re-Analysis of NGC 5506, GRS 1734-292 and MCG -05-23-16}
\label{reanalysis}
NGC 5506, GRS 1734-292 and MCG -05-23-16 have the most extreme values of coronal temperature of the sample. These values are obtained in literature using the \textsc{compTT} comptonization model \citep{titcomptt}. We check the kT$_e$ and $\tau$ values in spherical geometry of the corona for the above sources by re-analysing the \textit{NuSTAR} observations using the \textsc{nthcomp} model (\citealt{zdziarski1996} and \citealt{zycki1999}).\par 
NGC 5506 was observed with \textit{NuSTAR} (OBSID 60061323) on 2014 April 1. The observation was coordinated with the \textit{Swift} observatory (OBSID 00080413001), which observed the source, on 2012 April 2.In the re-analysis of NGC 5506, we fitted also the simultaneous \textit{Swift}/XRT data, but we did not re-extract the \textit{Swift}/XRT spectra.\par 
GRS 1734-292 was observed by NuSTAR on 2014 September 16 (OBSID 60061279002), for a total elapsed time of 43 ks.\par 
MCG -05-23-16 was observed on 2012 July 11–12 (OBSID 10002019), and on 2013 June 3–7 (OBSID 60001046). The first observation was conducted as a part of the \textit{NuSTAR} calibration campaign. The second observation was a science observation carried out simultaneously with a long \textit{Suzaku} observation. In our re-analysis we used only the \textit{NuSTAR} science observation. \par
First we reduced again the old \textit{NuSTAR} observations with the {\it NuSTAR} Data Analysis Software (NuSTARDAS) package (v. 1.6.0). Cleaned event files (level 2 data products) were produced and calibrated using standard filtering criteria with the {\sc nupipeline} task using the last new calibration files available from the {\it NuSTAR} calibration database (CALDB 20170120). The extraction radii of the circular region were 0.5 arcmin both for source and for background spectra, for all the sourced.\par
In their analysis of NGC 5506 \citet{matt05} found an X-ray spectrum composed by an absorbed power law (with $\Gamma \sim1.9$) with an exponential high-energy cutoff (E$_c=720^{+130}_{-190}$ keV), plus a moderately ionized reflection component, and ionized iron lines. They estimated the coronal parameters using the \textsc{compTT} comptonization model \citep{titcomptt}, and the \textsc{compps} model \citep{Poutanen}, founding, in the spherical geometry of the corona a coronal temperature of $440^{+230}_{-250}$ keV ($\sim 270$ keV) and an optical depth of 0.09 (0.14) respectively.\par 
\citet{tortosa17} found, for the \textit{NuSTAR} spectra of GRS 1734-292, a spectral shape of an absorbed power-law with photon index of $1.65$ and a very low exponential cutoff, $53^{+11}_{-8}$keV. They found a reflection fraction of $0.48 \pm 0.22$ and no evidence of relativistic features. Using the \textsc{compTT} model and assuming a spherical geometry for the Comptonizing corona, they fond a coronal temperature of $12.1^{+1.8}_{-1.3}$ keV and an optical depth $\tau=6.38^{+0.4}_{-0.5}$.\par
The analysis of the \textit{NuSTAR} spectrum of MCG -05-23-16, made up by \citet{balokovic15}, showed a primary power-law with an exponential high energy cutoff at $116^{+6}_{-5}$keV, a photon index of $1.85 \pm 0.01$ and the iron line with both narrow and broad component, the last one due to relativistic effects. \textsc{compTT} Comptonization model in the case of a spherical corona gives a coronal temperature kT$_e=25 \pm 2$ keV and a coronal optical depth $\tau = 3.5 \pm 0.2$.\par
We used models similar to the ones of \citet{matt05}, \citet{tortosa17} and \citet{balokovic15} to fit the NGC 5506, GRS 1734-292 and MCG -05-23-16 data, but we used the\textsc{relxillcp} and \textsc{xillver-comp} models (\citealt{xillver2010}; \citealt{relxill} and \citealt{dauser2014}) to model the relativistic or standard (respectively) reflection with the irradiation of the accretion by a power law with a \textsc{nthcomp}(\citealt{zdziarski1996} and \citealt{zycki1999}) Comptonization continuum.\par 
The values obtained with the re-analysis are showed in Table \ref{table:newpar}. The coronal optical depth values are extrapolated using the relation from \citet{belodorov1999}.
\begin{table}
\caption{List of some parameters obtained from the re-analysis of NGC 5506, GRS 1734-292 and MCG -05-23-16.}             
\label{table:newpar}      
\centering          
\begin{tabular}{ccccc}  
\hline\hline   
Source & $\Gamma$ & kT$_e$ (keV) & $\tau$ & $\chi^{2}/$d.o.f.\\
\hline
NGC 5506 & $1.73^{+0.09}_{-0.03}$& $400 \pm 200$ & $0.21$ & $1.1$ \\
GRS 1734-292 & $1.81 \pm 0.04$ & $16 \pm 3$ & $3.9$ & $1.02$ \\
MCG -05-23-16 & $1.93 \pm 0.01$ & $41 \pm 5$ & $1.73$ & $1.05$ \\
\hline
\hline                  
\end{tabular}
\end{table}
The values we found in our re-analysis are different from the literature values, especially the photon index $\Gamma$ (and so the optical depth). However the error bars on the coronal temperature almost are still the same. \par 
Even if the values we found are different from the literature ones, the $\tau$-kT$_e$ pairs follow the relation found previously with the literature values (see Figure \ref{fig:check}). 
\section{List of the sources}
\label{list}
\begin{itemize}
\item \textbf{NGC 5506} is a bright, nearby($z=0.006181$) Compton-thin \citep{wang1999}, narrow-line Seyfert 1 galaxy \citep{Nagar2002}. Its spectrum is well described by a power-law with $\Gamma=1.9 \pm.03$ with an high energy exponential cutoff at $720^{+130}_{-190}$  keV \citep{matt05}. NGC 5506 has a galactic absorption with a column density of $3.8 \times 10^{20} $cm$^{-2}$ \citep{LAB}. The observed 2-10 keV flux corrected for absorption is $6.2 \times 10^{-11}$ erg cm$^{-2}$s $^{-1}$ corresponding to a luminosity of $5.26 \times 10^{42}$ \lumcgs $\;$ \citep{matt05}.
\item \textbf{MCG -05-23-16}  is a nearby ($z=0.0085$, 36Mpc)Seyfert 1.9 galaxy (\citealt{veron1980}; \citealt{wegner2003}). This source has a complex structure of the fluorescent line emission, including both broad and narrow components produced by the disc and the torus reflection, respectively \citep{balokovic15}. It has an absorption with column density of $2.5 \times 10^{22}$cm$^{-2}$. The photon index of the primary power-law results to be $2.00 \pm 0.01$. It shows also an high energy cutoff at $116^{+6}_{-5}$keV \citep{balokovic15}
\item \textbf{SWIFT J2127.4+5654} ($z=0.0144$) is a narrow-line Seyfert 1. It was observed by \textit{NuSTAR} and \textit{XMM-Newton} in an observational campaign performed in November 2012. This source is affected only by the the Galactic column density absorption ($7.65 \times 10^{21}$ cm$^{-2}$, \citealt{LAB}). The primary emission of this source has a power-law spectral shape with a photon index of $2.08 \pm 0.01$ and a cutoff energy $E_c=108^{+11}_{-10}$ \citep{marinucci14a}.
\item  \textbf{IC4329A} is a nearby bright Seyfert galaxy ($z = 0.0161$ \citealt{willmer1991}; Galactic $N_H= 4.61 \times 10^{20}$ cm$^{-2}$, \citealt{LAB}). It has been observed by \textit{NuSTAR} quasi-continuously from 2012 August 12-16. The photon index of the primary power-law of IC4329A is $1.73 \pm 0.01$. The spectrum shows a cutoff at $184 \pm 14$ keV \citep{brenneman14}.
\item  \textbf{3C 390.3}  ($z = 0.056$) is a radio-loud Seyfert 1 galaxy with a weak reflection and a flat photon index. The timing properties of 3C390.3 do not differ from those of radio-quiet Seyferts \citep{gliozzi2009} and that there is no noticeable contribution from the jet to the X-ray emission \citep{sambruna2009}. It has a Galactic column density of $4 \times 10^{20}$cm$^{-2}$ \citep{LAB}, a photon index of the primary power-law of $1.70 \pm 0.01$   and a cutoff at energy of $116^{+24}_{-8}$ keV \citep{lohfink15}. 
\item  \textbf{3C 382} ($z=0.057870$) is a broad-line radio galaxy but its X-ray continuum is dominated by the Comptonizing corona similarly to radio-quiet Seyfert galaxies \citep{ballantyne14}. It has a Galactic absorption with a column density of $N_H=6.98 \times 10^{20}$cm$^{-2}$ \citep{LAB} and a weak, highly ionized warm absorber with $N_H \approx 1.4 \times 10^{21}$cm$^{-2}$ and $\log \xi =.5$, it has also a $\Gamma =1.68^{+0.03}_{-0.02}$ and a high energy cutoff at $214^{+147}_{-63}$ keV \citep{ballantyne14}.
\item \textbf{GRS 1734-292} ($z=0.0214$, corresponding to a distance of $87$ Mpc) is a Seyfert Galaxy originally discovered by the ART-P telescope aboard the \textit{GRANAT} satellite \citep{pavlinsky1992}. It has a total hydrogen column density in excess of $10^{22}$ cm$^{-2}$. The 2-10 keV flux for this source is $F_{\rm 2-10} =5.12^{+0.15}_{-0.08} \times 10^{-11}$ erg cm$^{-2}$ s$^{-1}$. GRS 1734-292 has the spectral slope of the primary power-law typical of a Seyfert galaxy in the \textit{NuSTAR} observation ($\Gamma\sim$1.65), with one of the lowest high energy cutoff ($53^{+11}_{-8}$ keV) measured so far by \textit{NuSTAR} \citep{tortosa17}.
\item \textbf{NGC 6814} ($z=0.0052$, \citealt{molina09}) is a Seyfert 1 Galaxy known to show X-ray variability by at least a factor of 10 over time scales of years \citep{mukai2003}. The $2-10$ keV absorption-corrected luminosities from the \textit{NuSTAR} observation is L$_{2-10}= 2.04 \times 10^{42}$ erg s$^{-1}$. It has a primary power-law with a photon index of $1.71^{+0.04}_{-0.03}$ and an exponential cutoff at $135^{+70}_{-35}$ keV \citep{tortosa17sub}.
\item \textbf{MCG 8-11-11} ($z=0.0204$) is a very X-ray bright AGN. The $2-10$ keV absorption-corrected luminosities from the \textit{NuSTAR} observation is L$_{2-10}= 5.13 \times 10^{43}$ erg s$^{-1}$. It has a primary power-law with a photon index of $1.77\pm0.04$ and an exponential cutoff at  $175^{+110}_{-50}$  keV \citep{tortosa17sub}.
\item \textbf{ Ark 564} ($z=0.02468$) is a narrow line Seyfert 1 galaxy. It has a steep X-ray spectrum, strong soft excess, and rapid variability. It is also extremely bright in the soft X-ray band ($F_{0.3-10 keV}=1.4 \times 10^{-10}$ erg cm$^{-2}$s $^{-1}$ \citep{kara2017}. Ark 564 has a photon index of $2.27 \pm 0.08$  and a very low cutoff energy value: $E_c=42 \pm 3$  \citep{kara2017}.
\item \textbf{PG 1247+267} is one of the most luminous known quasars at $z \sim 2$ and is a strongly super-Eddington accreting supermassive black hole (SMBH) candidate. It was observed by \textit{NuSTAR} in December 2014 for a total of 94 ks. From this observation it results that Pg 1247+267 has a primary power-law with a cutoff energy at $89^{+134}_{-34}$ keV and photon index of $2.35^{+0.09}_{-0.08}$ \citep{lanzuisi16}.
\item \textbf{Ark 120} ($z=0.033$) is a 'bare' Seyfert 1 galaxy, a system in which ionized, displaying neither intrinsic reddening in its IR continuum nor evidence for absorption in UV and X-rays absorption is absent (\citealt{matt14}, \citealt{porquet2017}). The spectrum of the source has a measure of the high energy cutoff value of $183^{+83}_{-43}$ keV \citet{porquet2017}. The photon index of the primary power-law is $1.87 \pm 0.02$ \citep{porquet2017}.
\item \textbf{NGC 7213} ($z = 0.005839$) is a low-luminosity active galactic nucleus that hosts a supermassive black hole of $\sim 10^8$ solar masses. It has also been classified as a low-ionization nuclear emission region galaxy (LINER) because of the low excitation observed in the narrow-line spectrum \citep{filippenko1984}. The photon index of the primary power-law of the spectrum of NGC 7213 is $1.84 \pm 0.03$. The sources does not have a cutoff measurements but shows only a lower limits on the cutoff energy of $E_c > 140$keV \citep{ursini15}.
\item \textbf{MCG 6-30-15} ($z=0.008$), is a Seyfert 1 galaxy with an extreme X-rays variability and a very broad Fe k$\alpha$ line emission, with an iron abundance  significantly higher than solar \citep{fabian2002}. Its primary power-law show a photon index of $2.06 \pm 0.01 $ and a lower limit on the high enery cutoff which results to be $>110$ keV \citep{marinucci14b}. 
\item \textbf{NGC 2110} ($z=0.008$) is a bright Seyfert 2 galaxy. it shows a prominent Fe k$\alpha$ line with a variable intrinsic emission and shows a cutoff energy $E_c > 210$ keV, with no detectable contribution from Compton reflection \citep{marinucci15}. The source has several layers of absorbing material with column densities in the range $2-6 \times 10^{22}$cm$^{-2}$ \citep{rivers2014}.
\item  \textbf{Mrk 335} ($z=0.0257$) is a arrow-line Seyfert 1 galaxy that show narrower broad emission-line components than typical Type 1 AGN \citep{grier12}. It has been observed by \textit{NuSTAR} in June and July 2013. The Galactic absorption for this source is of $3.56 \times 10^{2}$ cm$^{-2}$ \citep{LAB}. Its primary power-law spectrum shows a photon index of  $2.14^{+0.02}_{-0.04}$ and a cutoff energy $E_c>174$ \citep{parker2014}.
\item \textbf{Fairall 9} ($z=0.047016$) is a Seyfert 1 galaxy. It has been observed by \textit{NuSTAR} in May, 2014 and does not show any significant absorption other than Galactic \citep{lohfink16}. The photon index of its primary power-law is $1.96^{+0.01}_{-0.02}$, which shows a cutoff $E_c>242$ \citep{lohfink16}.
\item \textbf{Mrk 766} ($z=0.012929$) is a narrow line Seyfert 1 galaxy which shows spectral variability in the X-rays \citep{risaliti2011}. Its X-ray spectrum is well fitted by a power-law with photon index  $2.22^{+0.02}_{-0.03}$ and an exponential cutoff with a lower limit of $>441$keV \citep{buisson17}.
 \item \textbf{PG 1211+143} ($z=0.080900$) is a bright radio-quiet quasar which belongs to the class of Narrow Line Seyfert 1 galaxies and presents an archetypical case for the ultra-fast outflows. The amount of Galactic neutral absorption along the line of sight is  $2.7 \times 10^{20}$ cm$^{-2}$ \citep{LAB}. The photon index of the primary power-law of the spectrum of this source is $2.51 \pm0.2$ with a lower limit on the exponential cutoff of $>124$ keV \citep{zoghbi15}.
 \end{itemize}
\end{appendix}
\end{document}